\documentclass[%
 twocolumn,
showpacs,preprintnumbers,
 amsmath,amssymb,
pra,
]{revtex4-1}
\usepackage[utf8]{inputenc}	
\usepackage[T1]{fontenc}	
\usepackage{graphicx}		
\usepackage{dcolumn}		
\usepackage{booktabs}		
\usepackage{mathtools}		
\usepackage{amsmath,array}
\usepackage{hyperref}

\usepackage{relsize}
\usepackage{enumerate}
\usepackage{float}
\usepackage{braket}

\usepackage[FIGTOPCAP]{subfigure}

\usepackage[squaren]{SIunits}	
\usepackage {natbib}
\usepackage{pdfpages}
\usepackage{enumitem}
\let\originaleqref\eqref
\renewcommand\figurename{Fig.}
\renewcommand{\eqref}{Eq.~\originaleqref}

\newcommand{\fref}[1]{\figurename~\ref{#1}}

\newcommand{\ttt}{(\tau,\theta)} 

\newcommand{\pdiff}[2]{\frac{\partial #1}{\partial #2}}

\graphicspath{{Billeder/}}

\begin{document}
\title{Parameter estimation by multi-channel photon counting}
\author{Alexander Holm Kiilerich}
\email{kiilerich@phys.au.dk}
\author{Klaus Mølmer}

\affiliation{Department of Physics and Astronomy, Aarhus University, Ny Munkegade 120, DK 8000 Aarhus C. Denmark}
\date{\today}
\bigskip

\begin{abstract} 
The physical parameters governing the dynamics of a light emitting quantum system can be estimated from the photon counting signal. The information available in the full detection record can be analysed by means of the distribution of waiting times between detection events. Our theory  allows calculation of the asymptotic, long time behaviour of the sensitivity limit, and it applies to emission processes with branching towards different final states accompanied by the emission of distinguishable photons. We illustrate the theory by application to a laser driven $\Lambda$-type atom.
\end{abstract}

\maketitle
\noindent
\section{introduction}

Atoms and atom-like systems with discrete energy states are widely used for precision measurements of time and frequency and as sensitive probes of fields or other influences on the system behaviour. The random character of measurements on quantum systems fundamentally limits the information achievable, but quantum states with squeezed uncertainty of particular observables, and entangled states of multi-particle systems have been identified as particularly sensitive initial states for (repeated) single-shot experiments, see e.g. \cite{Giovannetti19112004,review}.

Rather than many repeated experiments we have the situation in mind of a single quantum system probed continuously over time. One must then take the measurement back action into account at all measurement steps, and this is conveniently done in the quantum trajectory formalism. This provides, conditioned on the measurement record \cite{Mabuchi1996,PhysRevA.64.042105,likelihood}, both the state of the quantum system and, via Bayes' rule, the probabilities of different candidate values of the estimated parameter. If the system is subject to damping and decoherence, and behaves in an ergodic manner, one may regard data obtained at sufficiently well separated moments of time as statistically independent. Continuous probing of the same system for a long time $T$ can hence be thought of as a number of $N$ independent experiments with $N\propto T$, and we expect an estimation error scaling asymptotically as $1/\sqrt{T}$.

To confirm this expectation and to identify the quantitative performance of continuous probing we shall address the Cramér-Rao bound (CRB) \cite{Cramer},
\begin{equation}
\label{eq:CRB}
[\Delta S(\theta)]^2 \geq \frac{1}{F(\theta)},
\end{equation}
which expresses the lower limit of the statistical variance $[\Delta S(\theta)]^2$ of any unbiased estimator for an unknown quantity $\theta$ by the Fisher information,
\begin{equation}
\label{Fisher}
F(\theta)=-\sum_D\frac{\partial^2 \ln L(D|\theta)}{\partial \theta^2}L(D|\theta)
\end{equation}
where $L(D|\theta)$ in \eqref{Fisher} is the likelihood to obtain measurement data $D$ conditioned on the value $\theta$.
For $n_{\textrm{rep}}$ repeated experiments, \eqref{eq:CRB} is written with an extra factor $1/n_{\textrm{rep}}$, and the bound applies in the limit of $n_{\textrm{\textmd{rep}}} \gg 1$. In our case, however, $D$ represents a single, time-dependent detection record, and the asymptotic convergence of our estimate should follow from the probing time dependence of $F(\theta)$ in the long time limit.

\begin{figure}
\centering

\includegraphics[width=0.9\columnwidth]{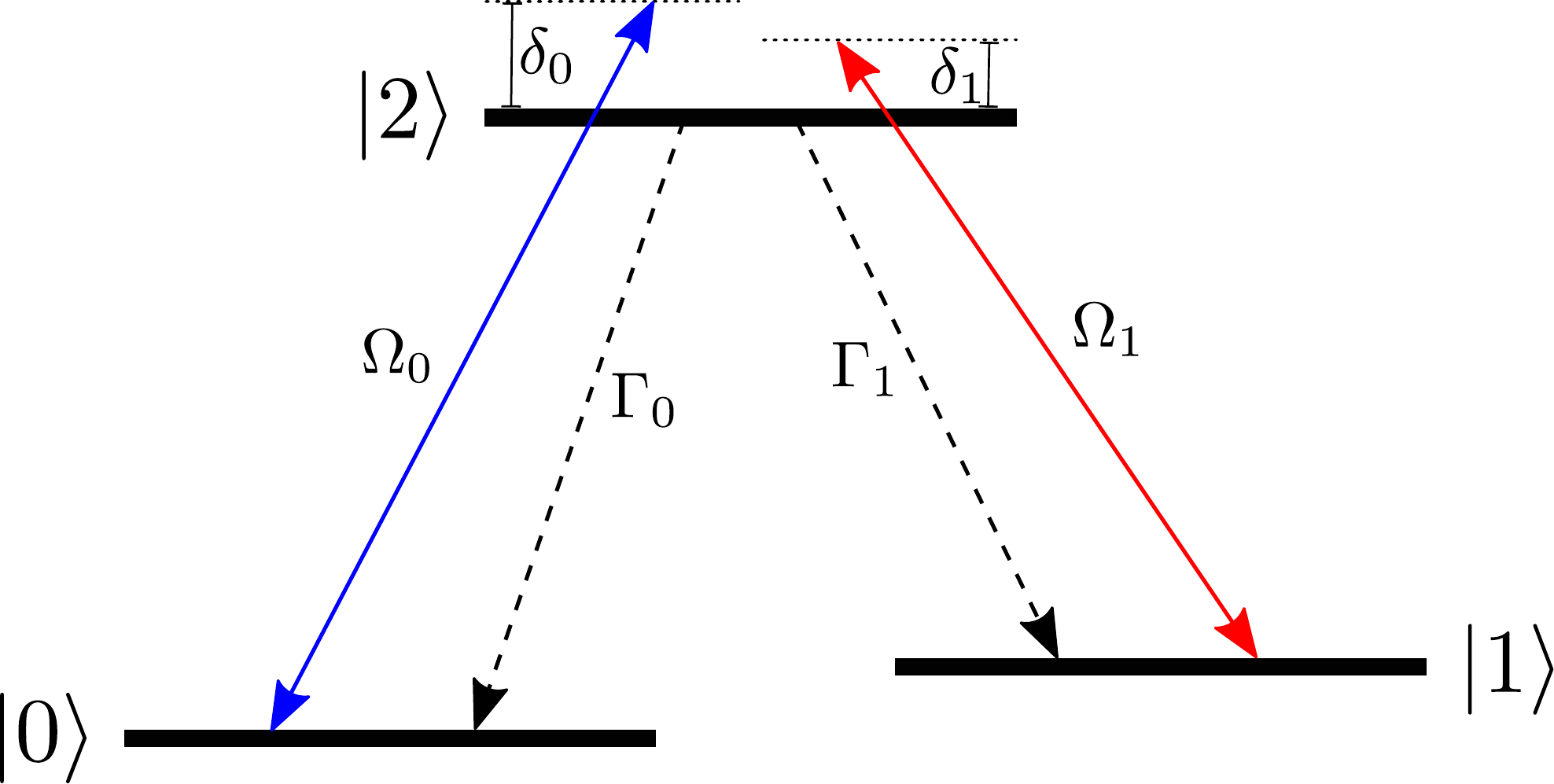}
\caption{\textsl{(Color online) A quantum $\Lambda$-system with laser driven $\ket{0}\leftrightarrow\ket{2}$ and $\ket{1}\leftrightarrow\ket{2}$ transitions with Rabi frequencies $\Omega_0$  and $\Omega_1$, detunings $\delta_0$ and $\delta_1$ and excited state decay rates $\Gamma_{0}$ and $\Gamma_{1}$.}}
\label{fig:lambda} 
\end{figure}

In this article we consider detection by photon counting of the radiation emitted by a quantum light source. For a closed two-level transition, the discrete waiting times between detection events form independent and identically distributed stochastic variables, and we have previously shown \cite{delay} that this simplifies the evaluation of the Fisher information and the Cramér-Rao bound. Here, we generalize the approach of \cite{delay} to the case of multi-level systems with distinguishable emission processes and branching of the decay towards different final states. This situation is exemplified by the $\Lambda$-system depicted in \fref{fig:lambda}, with an excited state from which spontaneous decay occurs towards two different ground states.
Since the decay processes leave the atom in different states, subsequent time intervals between detector clicks are not independent. The purpose of this article is to derive a theory that allows calculation of the Fisher information and the Cramér-Rao bound for parameter estimation with two-channel (and more general multi-channel) counting signals.


The article is outlined as follows.  In Sec. II, we discuss how single-channel and multi-channel photon counting records can be reorganized as the sampling of uncorrelated stochastic variables and how the Fisher information can be calculated from the distribution of waiting times between detections in different channels.  In Sec. \ref{sec:waiting}, we present a master equation analysis of the theoretical waiting time distribution functions that allow practical calculation of the Fisher information. In Sec. IV, we present the filter function that should be applied to multi-channel measurement data records to achieve parameter estimates that reach the Cramér-Rao bound. In Sec. V we show results for the $\Lambda$-system in \fref{fig:lambda}, and in Sec. VI, we conclude the analysis.

\section{Single-channel and multi-channel counting signals} \label{counting}

A photon counting detection record contains the discrete times of detection events $D=\{t_k\}$, and, if the emitter always jumps to the same state when a photon is detected, measurement intervals $\tau_k=t_{k+1}-t_k$  between detector clicks are independent and identically distributed stochastic variables. A data record with $N+1$ count events, thus, yields $N$ independent samples of the waiting time probability distribution $w(\tau)$. Each registered waiting time $\tau_k$ falls within a short interval $[\tau^{[i]}-\Delta \tau/2,\tau^{[i]}+\Delta \tau/2]$ with probability $w_i=w(\tau^{[i]})\Delta \tau$, and the data record $D$, fully represented by the set of numbers $n_i$ of registered waiting times in all intervals, is statistically governed by a multinomial distribution, $(\sum_i n_i)!\prod_iw_i^{n_i}/\prod_i n_i!$.

For a given total probing time $T$, the total number of registered intervals, $N= \sum_i n_i$ is itself a stochastic variable, governed by a probability distribution $P_N$, and hence the likelihood for the data record $D=\{n_i\}$ is
\begin{align}\label{eq:likelihood}
L(D|\theta) = \frac{\left(\sum_i n_i\right)!\prod_iw_i^{n_i}}{\prod_i n_i!}P_{N=\sum_i n_i},
\end{align}

The conditional dependence on the quantity $\theta$ stems from the $\theta$-dependence of the $w_i$'s and $P_N$ in Eq. (\ref{eq:likelihood}), and  the Fisher information \eqref{Fisher} can be evaluated directly,
\begin{align}\label{F_full}
F(\theta) = \overline{N} \sum_i\frac{1}{w_i}\left(\frac{\partial w_i}{\partial \theta}\right)^2
+
\sum_N \frac{1}{P_N} \left(\frac{\partial P_N}{\partial \theta}\right)^2,
\end{align}
where $\overline{N}$ denotes the mean value of $N$.

It is convenient to rearrange the terms in Eq. (\ref{F_full}) in two different contributions,
\begin{align} \label{eq-non-poisson}
F(\theta) &= F_{\text{Poisson}}(\theta)+F_N(\theta).
\end{align}
The first term
\begin{align}\label{eq:Fisher-single}
F_{\text{Poisson}}(\theta) \equiv \sum_i \frac{1}{\overline{n}_i} \left(\frac{\partial \overline{n}_{i}}{\partial \theta}\right)^2 =  \int \frac{1}{\overline{n}(\tau)} \left(\frac{\partial \overline{n}(\tau)}{\partial \theta}\right)^2 \mathrm{d}\tau,
\end{align}
reflects the similarity between the multinomial distribution and the Poisson distribution for each $n_i\ll N$ with $\overline{n}_i(\theta) = \overline{N}(\theta)w_i(\theta)$. In the last  step, we have  transformed the sum over time intervals into an integral with $\overline{n}_i(\theta)= \overline{n}(\tau,\theta)\Delta \tau$.
See, e.g., \cite{CRB,mereCRB} for similar arguments applied to high-resolution spatial measurements by scattering of coherent light or to probing of the motion of a Bose condensate.

\eqref{eq:Fisher-single} is, indeed, the Fisher information for uncorrelated Poisson distributed variables $n_i$, resulting in a sum, $N=\sum_i n_i$, which is also Poisson distributed. The total number of photons emitted from quantum light sources may, however, show sub- or super-Poissonian counting statistics \cite{RevModPhys.54.1061}, and the second term in (\ref{eq-non-poisson}),
\begin{align}\label{eq:corr}
F_N(\theta) \equiv \sum_N \frac{1}{P_N(\theta)} \left(\frac{\partial P_N(\theta)}{\partial \theta}\right)^2-\frac{1}{\overline{N}(\theta)}\left(\frac{\partial \overline{N}(\theta)}{\partial \theta}\right)^2
\end{align}
accounts for the deviation of the information held by the true statistics $P_N$ from that of a Poisson distribution.
The two expressions (\ref{F_full},\ref{eq-non-poisson}) are easily proven to be identical (note that a term with mixed derivatives vanishes because of the independence on $\theta$ of the integral of $w(\tau)$ over time).

The waiting times are identically distributed random variables, and the stochastic counting process is a renewal process \cite{Cox}. We are interested in systems with no dark steady states, i.e., the fluorescence is persistent and the waiting time distribution does not have long tails. In the asymptotic limit of large $T$, where the CRB applies, a Central Limit Theorem for such renewal processes ensures that the distribution $P_N$ converges asymptotically to a normal distribution with a mean value $\overline{N}$ and a variance $V \equiv \mathrm{Var}(N)$ which are both proportional with $T$ \cite{Feller}. In this limit we can thus evaluate
\eqref{eq:corr},
\begin{align}\label{eq:Fisher_normal}
F_N(\theta)= \left(
\frac{1}{V(\theta)}- \frac{1}{\overline{N}(\theta)}\right) \left(\frac{\partial \overline{N}(\theta)}{\partial \theta}\right)^2.
\end{align}
As easily understood, the correction (\ref{eq:Fisher_normal}) to the Fisher information is positive(negative) if the total number $N$ fluctuates less(more) than the Poisson distribution.

Ref. \cite{delay} did not take the correction (\ref{eq:Fisher_normal}) into account. This was justified by the focus in that article on a saturated transition, where the distribution of waiting times yields much more information than the total count, and where the first term in (\ref{eq-non-poisson}) therefore completely dominates the Fisher information. The relative significance of the terms in \eqref{eq-non-poisson} depends on the physical system, and in the examples studied in Sec. VI in this article, the full count statistics term cannot be ignored.

Since $\overline{N}$, $V$ and $\overline{n}(\tau,\theta)$ are all proportional to the probing time $T$, also the Fisher information is proportional to $T$, and we conclude from \eqref{eq:CRB} that the estimation error decreases asymptotically as $\sim 1/\sqrt{T}$.

To evaluate our expression for the Fisher information (\ref{eq-non-poisson}), we need to determine how the waiting time distribution $\overline{n}(\tau)$ and the ensuing $\overline{N}$ and $V$ depend on $\theta$. This information can be retrieved from the system master equation, but let us first turn to the more general case of signals from quantum emitters observed by photon counters that distinguish between different decay channels, \textit{e.g.}, by making use of the polarization or frequency of the emitted photons.

For generality we assume that there are $M$ such channels (for the $\Lambda$-system in \fref{fig:lambda}, $M=2$). Our analysis is restricted to the case for which detection of a photon in channel $m$ accompanies a jump of the emitter into a definite state $|\phi_m\rangle$, which is the initial state for the subsequent evolution of the system. This is not a requirement for the Bayesian analysis, but our calculation of the Fisher information relies on definite waiting time distributions after detection in each channel. These waiting time distributions until the next detection event, thus, depend on $m$, the channel of the most recently detected photon, and we can sort the detection record into lists $\{\tau_k\}_{mm'}$ containing the duration of time intervals between detection in channel $m$ followed by subsequent detection in channel $m'$.   These lists, in turn, sample the corresponding waiting time distributions in an independent and uncorrelated manner, and, for our parameter estimation, they retain all the information available in the multi-channel detection record.

The combinations $mm'$ define $M^2$ interval types, and for each $mm'$, the number $n_{mm',i}$ denotes the number of waiting times $\tau$ registered in intervals $[\tau^{[i]}-\Delta \tau/2,\tau^{[i]}+\Delta \tau/2]$.
The likelihood function in \eqref{Fisher} now factorizes as a product of weighted multinomial distributions,
\begin{align}
L(D|\theta)= \prod_{mm'}L_{mm'}(D|\theta),
\end{align}
where the likelihood for each type, $L_{mm'}(D|\theta)$, is as given in \eqref{eq:likelihood}, and the single channel result (\ref{eq-non-poisson}) is readily generalized.

In particular,
\begin{equation}
\label{F}
F_{\text{Poisson}}(\theta)=\sum_{mm'} \int \frac{1}{\overline{n}_{mm'}(\tau,\theta)}\left(\frac{\partial \overline{n}_{mm'}(\tau,\theta)}{\partial \theta}\right)^2 \mathrm{d}\tau,
\end{equation}
where $\overline{n}_{mm'}(\tau,\theta)$ is the theoretically expected distribution of intervals of type $mm'$ and duration $\tau$. The correction due to the count statistics with mean value $\overline{N}_m$ but a non-Poissonian variance $V_m$ in each channel is in the asymptotic limit given by
\begin{align}\label{eq:F_Nfinal}
F_N(\theta)=\sum_m\left(\frac{1}{V_m}-\frac{1}{\overline{N}_m}\right) \left(\frac{\partial \overline{N}_m}{\partial \theta}\right)^2.
\end{align}

\section{Waiting time distributions } \label{sec:waiting}

We obtain the distribution functions $w_{mm'}(\tau,\theta)$ and $\overline{n}_{mm'}(\tau,\theta)$ by solving effective master equations where the unknown quantity $\theta$ is one of the Hamiltonian or damping parameters. With the understanding that our results may be finally evaluated and varied with respect to the parameter of interest, we suppress, in this section, the variable $\theta$ from the equations.

The average behaviour of an atomic  quantum system decaying by spontaneous emission of photons into broad-band photon reservoirs is described by a master equation of the form $(\hbar=1)$ \cite{WM},
\begin{align}
\frac{\mathrm{d}\rho}{dt}=-i[\hat{H}_0,\rho]+
\sum_m \left( \hat{C}_m \rho \hat{C}_m^\dagger-\frac{1}{2}\{\hat{C}_m^\dagger\hat{C}_m,\rho\} \right),
\label{eq:master0}
\end{align}
where the operators $\hat{C}_m$ represent jump processes in the atom associated with decay and emission of different, distinguishable kinds of radiation. While decay processes may preserve, e.g., coherences between excited Zeeman states in the ground state after the emission of light of linear or circular polarization, we emphasize that our analysis of the Fisher information is restricted to the case in which a jump $\hat{C}_m$  puts the system in a definite final state $|\phi_m\rangle$, from which the dynamics proceeds. This is for example the case for the three level atom, shown in \fref{fig:lambda}, where the two operators, $\hat{C}_0 = \sqrt{\Gamma_{0}}|0\rangle \langle 2|$ and  $\hat{C}_1 = \sqrt{\Gamma_{1}}|1\rangle \langle 2|$ describe decay into the ground states $|0\rangle$ and $|1\rangle$ with rates $\Gamma_{0}$ and $\Gamma_{1}$, respectively.

With the interpretation of quantum trajectories or Monte Carlo wave functions \cite{delay,CARMICHAEL,MCWF} as the states of dissipative quantum systems conditioned on the outcome of continuous probing of their emitted radiation, it is possible to simulate realistic detection records. The jumps into state $|\phi_m\rangle$ are governed by the rate $\langle \hat{C}_m^\dagger \hat{C}_m\rangle$ where the expectation value is calculated as function of time for a given evolving wave function. On average, the stochastically evolving wave functions reproduce the master equation and therefore the average number of these jumps equals the value obtained by the density matrix describing the un-observed quantum system. For probing over long times $T$, we thus get the average number of jumps into state $|\phi_m\rangle$, $\overline{N}_m = \textrm{Tr}(\hat{C}_m^\dagger \hat{C}_m\rho^{\textrm{st}})T$, where $\rho^{\textrm{st}}$ is the steady state density matrix solution to the master equation (\ref{eq:master0}).

For the distributions of intervals between detector clicks we now have $\overline{n}_{mm'}(\tau)=\overline{N}_m w_{mm'}(\tau)$, where $w_{mm'}(\tau)\mathrm{d}\tau$, is the probability that after a jump into $|\phi_m\rangle$, the next emission event is detected in channel $m'$ in $[\tau,\tau + \mathrm{d}\tau]$. To determine the function $w_{mm'}(\tau)$, we note that the terms $\sum_m \hat{C}_m \rho \hat{C}_m^\dagger$ in \eqref{eq:master0} account for the feeding of the system ground states associated with the emission process, \textit{i.e.}, they describe terms in the reduced system density matrix, correlated with single-photon excited states of the modes of the radiation field. If the system has just been put into the state $|\phi_m\rangle$ due to detection of a photon in channel $m$, the probability that no photon is detected until a certain later time $\tau$ is equal to the population of the zero-photon component of the combined state of the system and the environment at that time. This is, in turn, given by the trace of the un-normalized density matrix, $\tilde{\rho}$, which evolves from the initial state $\tilde{\rho}|_m(\tau=0) = |\phi_m\rangle\langle\phi_m|$, omitting the ground state feeding term of the master equation,
\begin{align}
\frac{\mathrm{d}\tilde{\rho}}{\mathrm{d}t}=-i[\hat{H}_0,\tilde{\rho}]
-\frac{1}{2}\sum_m\{\hat{C}_m^\dagger\hat{C}_m,\tilde{\rho}\}.
\label{eq:mastertilde}
\end{align}
The resulting $\tilde{\rho}|_m(\tau)$ is equivalent to the so-called no-jump wave function \cite{MCWF} evolving from the state $|\phi_m\rangle$ by the non-hermitian Hamiltonian $\hat{H}_{\textrm{eff}} =  \hat{H}_0 - \frac{i}{2} \sum_m \hat{C}_m^\dagger\hat{C}_m$.
The probability $w_{mm'}(\tau)\mathrm{d}\tau$ that after a detector click at time $t$ of type $m$, the next click is of type $m'$ and occurs in the time interval $[t+\tau,t+\tau+\mathrm{d}\tau]$, is now given by
\begin{equation}
w_{mm'}(\tau)\mathrm{d}\tau = \textrm{Tr}(\hat{C}_{m'}^\dagger\hat{C}_{m'}\tilde{\rho}(\tau))\mathrm{d}\tau.
\end{equation}
It follows from the master equation that these waiting time distributions are normalized according to
\begin{align}\label{eq:w_norm}
\sum_{m'} \int_0^\infty w_{mm'}(\tau)\mathrm{d}\tau = 1.
\end{align}
With the values thus found theoretically for $\overline{N}_m$ and  $w_{mm'}(\tau)$, we know $\overline{n}_{mm'}(\tau)$, and we can evaluate the Fisher information in (\ref{F}).

If photons are detected with only finite efficiency $\eta$, this is equivalent to a fraction $1-\eta$ of the quantum jumps passing unnoticed. The corresponding un-normalized state $\tilde{\rho}$ conditioned on no detection events is then found by including a ground state feeding term, $(1-\eta)\hat{C} \tilde{\rho} \hat{C}^\dagger$, in the no-jump master equation to account for the unobserved emission \cite{delay}. In the multi-channel case, if different channels are monitored with detector efficiencies $\eta_m$, we obtain the no-detected-jump master equation
\begin{align}
\frac{\mathrm{d}\tilde{\rho}}{\mathrm{d}t}=-i[\hat{H}_0,\rho]+
\sum_m \left( (1-\eta_m)\hat{C}_m \rho \hat{C}_m^\dagger-\frac{1}{2}\{\hat{C}_m^\dagger\hat{C}_m,\tilde{\rho}\} \right).
\label{eq:mastertildeeta}
\end{align}
The solutions of this equation for initial states $\tilde{\rho}|_{m}(\tau=0) = |\phi_m\rangle\langle\phi_m|$ yield the waiting time distributions between the \textit{detected} emission events \cite{delay},
\begin{equation}
w_{mm'}(\tau)\mathrm{d}\tau = \eta_{m'}\textrm{Tr}(\hat{C}_{m'}^\dagger\hat{C}_{m'}\tilde{\rho}|_{m}(\tau))\mathrm{d}\tau,
\label{w_eta}
\end{equation}
which are normalized as in \eqref{eq:w_norm}. The average number of detected events in channel $m$ during probing for time $T$ is $\overline{N}_m = \eta_m \textrm{Tr}(\hat{C}_m^\dagger \hat{C}_m\rho^{\textrm{st}})T$, and with the resulting $\overline{n}_{mm'}(\tau)=\overline{N}_m w_{mm'}(\tau)\mathrm{d}\tau$, we can calculate the Fisher information according to (\ref{F}).

The Fisher information \eqref{eq-non-poisson} also depends on the moments of the total count statistics (\ref{eq:F_Nfinal}). Calculating the variance in the photon count is one of the founding problems of quantum optics \cite{RevModPhys.54.1061}, and we give here a simple recipe relying on quantities already derived.
Consider the duration $T_N=\sum_{i=1}^{N} \tau_i$ of $N$ waiting time intervals. $T_N$ has a mean value $\overline{T}_N=N\overline{\tau}$ and a variance $\textrm{Var}(T_N)=N \textrm{Var}(\tau)$ The corresponding uncertainty in the number of detection events in a definite time interval follows, $\sqrt{\textrm{Var}(N)} = (\mathrm{d} N/\mathrm{d}T_N)\times \sqrt{\textrm{Var}(T_N)} = \sqrt{N}\sqrt{\textrm{Var}(\tau)}/\overline{\tau}$  -  where, for an exponential waiting time distribution  with $\textrm{Var}(\tau)=\overline{\tau}^2$, we recover the Poissonian statistics.

The quantities $\overline{\tau}$ and $\textrm{Var}(\tau)$ can be evaluated from the waiting time distribution functions, and in the multi-channel case, the $k$'th moment of $\tau$ pertaining to the channel $m$ is given as
\begin{align}
(\overline{\tau^k})_m = \int \tau^k w_{mm}(\tau)\, \mathrm{d}\tau,
\end{align}
where $w_{mm}(\tau)$ is the distribution function for waiting times between photo detection events in the channel $m$, and is obtained by solving \eqref{eq:mastertildeeta} with efficiencies $\eta_m$ and $\eta_{m'\neq m}=0$.
One may then calculate
\begin{align}\label{eq:V_m}
V_m = \frac{\textrm{Var}(\tau)_m}{\overline{\tau}_m^2}\overline{N}_m,
\end{align}
clearly identifying whether $N_m$ follows sub- or super-Possonian statistics.

\section{Achieving the Cramér-Rao bound}

The CRB concerns the asymptotic sensitivity, and we assume that the value of $\theta$ is already known to within a small error $\delta \theta$ from an offset value which, for convenience, we redefine as $\theta=0$.
For single channel Poisson distributed counting signals, a simple linear filter achieves the CRB \cite{CRB,delay,gammelmark} and motivates an ansatz for the multi-channel estimator, when the total counts in each channel are Poisson distributed, of the form,
\begin{equation}
S_{P}(n_{mm'}(\tau))=\sum_{mm'}\left(\int g_{mm'}(\tau)n_{mm'}(\tau) \, \mathrm{d}\tau
+C_{mm'}\right),
\label{S}
\end{equation}
which weighs the actual recorded distributions of waiting times $n_{mm'}(\tau)$  with gain functions $g_{mm'}(\tau)$ and constant offsets $C_{mm'}$, chosen to ensure the correct mean value and to minimize the statistical variance of the estimator.

We assume that $\delta\theta$ is sufficiently small that the corresponding change in the expected waiting time distribution $\overline{n}_{mm'}(\tau,\delta\theta)$ in (\ref{S}) is well represented by a first order Taylor expansion.  To cancel the zeroth order terms in (\ref{S}), we then pick
\begin{align}
C_{mm'}=- \int g_{mm'}(\tau)\overline{n}_{mm'}(\tau,0) \, \mathrm{d}\tau,
\end{align}
and for data in complete accordance (no noise) with the expected mean, we obtain to first order
\begin{equation}
\nonumber
S_P(\overline{n}_{mm'}(\tau,\delta\theta))=\delta \theta\sum_{mm'} \int g_{mm'}(\tau) \left.\pdiff{\overline{n}_{mm'}(\tau,\theta)}{\theta}\right|_{\theta=0} \, \mathrm{d}\tau.
\label{smiddel_flere}
\end{equation}

The uncorrelated, Poisson distributed count signals allow calculation of the variance of the estimator (\ref{S}),
\begin{align}\label{eq:esti_var}
(\Delta S_P)^2=\sum_{mm'}\int g^2_{mm'}(\tau)\overline{n}_{mm'}(\tau,0) \, \mathrm{d}\tau.
\end{align}
Next, the signal-to-noise ratio, $$(\mathrm{SNR})^2\equiv \frac{ S_P^2(\overline{n}_{mm'}(\tau,\delta \theta))}{(\Delta S_P)^2},$$ can be maximized by the Cauchy Schwarz inequality, $|\langle v_k(\tau),u_k(\tau)\rangle|^2\leq \langle v_k(\tau),v_k(\tau)\rangle\langle u_k(\tau),u_k(\tau)\rangle$, where $u_k(\tau)$ and $v_k(\tau)$ are functions of the continuous variable $\tau$ and the discrete variable $k=(mm')$.

Applying the inequality with: $v_k(\tau)=\left.\delta   \theta  \pdiff{\overline{n}_{mm'}\ttt}{\theta}\right|_{\theta=0}\overline{n}_{mm'}^{-1/2}(\tau,0)$ and $u_k(\tau)=g_{mm'}(\tau)\overline{n}_{mm'}^{1/2}(\tau,0)$, we obtain
\begin{align}
(\mathrm{SNR})^2\leq (\delta \theta)^2F_{\text{Poisson}}(\theta)
\label{SNR}
\end{align}
with $F_{\text{Poisson}}(\theta)$ given in Eq. (\ref{F}). The Cauchy Schwarz inequality is saturated when the functions $v_k(\tau)$'s and $u_k(\tau)$'s are proportional, which occurs when
\begin{align}
g_{mm'}(\tau)=\frac{\beta}{\overline{n}_{mm'}(\tau,0)}\left.\pdiff{\overline{n}_{mm'}\ttt}{\theta} \right|_{\theta=0},
\label{gain_i}
\end{align}
where the constant $\beta$ is the same for all $mm'$.

The requirement that data in complete accordance with the expected distributions $\overline{n}_{mm'}\ttt$ should lead to $S( \overline{n}_{mm'}(\tau,\delta\theta) )=\delta\theta$ establishes that, in fact, $\beta$ must be the inverse Fisher information
$
\beta=F_{\text{Poisson}}^{-1}(\theta).
$

The shot noise limit, $\mathrm{SNR}=1$ in (\ref{SNR}), defines the lowest distinguishable value of $\delta\theta = 1/\sqrt{F_P(\theta)}$, and collecting the results provides the linear estimator \eqref{S} in terms of the expected and the actually measured distribution of time intervals between the detector clicks,
\begin{align}
\nonumber
S_P(n_{mm'}(\tau))&=F_{\text{Poisson}}^{-1}(\theta)\sum_{mm'} \int \left.\pdiff{\overline{n}_{mm'}\ttt}{\theta}\right|_{\theta=0}
\\
&\quad\times
\left(\frac{n_{mm'}(\tau)}{\overline{n}_{mm'}(\tau,\theta)}-1\right) \, \mathrm{d}\tau.
\label{estimator flere}
\end{align}
The prior estimate is adjusted according to the discrepancy between the recorded waiting times and those expected from that prior. The Fisher information appears as a normalizing factor which reflects that larger adjustments may apply when the uncertainty is large.
Still, we recall that this expression only applies asymptotically and that it is valid only if the first order Taylor expansions in the deviation from our prior guess are accurate enough, see also \cite{GutaGill}.

In the general case of non-Poissonian counting statistics $V_m\neq \overline{N}_m$, and in the derivation above we must explicitly treat the $N_m$'s as independent stochastic variables that can themselves have a $\theta$-dependence. Factorizing the waiting time distributions $\overline{n}_{mm'}(\tau)=\overline{N}_m(\theta) w_{mm'}(\tau,\theta)$ allows us to employ separate gains for each $N_m$ in \eqref{S}, and when the variance of the estimator \eqref{eq:esti_var} is corrected to include the proper variances $V_m$, the arguments given in this section carries over and the estimator acquires an extra term depending on the photon counts $N_m$,
\begin{align} \label{general:S}
\nonumber
S(n_{mm'}(\tau))&=F^{-1}(\theta)\bigg[\sum_{mm'} \int \left.\pdiff{\overline{n}_{mm'}\ttt}{\theta}\right|_{\theta=0}
\\
&\quad\times
\left(\frac{n_{mm'}(\tau)}{\overline{n}_{mm'}(\tau,\theta)}-1\right) \, \mathrm{d}\tau
\nonumber
\\
&\quad+
\sum_m  \left.\pdiff{\overline{N}_{m}\ttt}{\theta}\right|_{\theta=0}
\left(N_m-\overline{N}_m(\theta)\right)
\nonumber
\\
&\quad\times
\left(
\frac{1}{V_m(\theta)}- \frac{1}{\overline{N}_m(\theta)}\right)\bigg].
\end{align}
The Fisher information is given in \eqref{eq-non-poisson}, and \eqref{general:S} constitutes a linear estimator that exhausts the information in the multi-channel photon counting data record and, hence, achieves the Cramér-Rao Bound asymptotically.

\section{Photon counting from a laser driven $\Lambda$-type atom}

As an example, we apply the formalism to a $\Lambda$-type system coupled to two laser fields, as shown in \fref{fig:lambda}. The couplings are described by Rabi frequencies $\Omega_0$ and $\Omega_1$ and laser-atom detunings $\delta_0$ and $\delta_1$  as indicated in the figure. We assume no direct coupling between $\ket{0}$ and $\ket{1}$, and that the decay into these two ground states is distinguishable, either by the polarization or by well-separated frequencies of the emitted photons.

In the rotating wave approximation, the Hamiltonian of the system can be written in matrix form as ($\hbar=1$),
\begin{align}
\hat{H}_0=\left( \begin{array}{ccc}
\delta_0 & 0 & \frac{\Omega_0}{2} \\
0 & \delta_1 & \frac{\Omega_1}{2}\\
\frac{\Omega_0}{2} & \frac{\Omega_1}{2} & 0\\
\end{array} \right).
\label{hamilton lambda}
\end{align}
The decay from $\ket{2}$ to $\ket{0}$ with rate $\Gamma_{0}$ and from $\ket{2}$ to $\ket{1}$ with rate $\Gamma_{1}$ (\fref{fig:lambda}) lead to a measurement record of photo detection events, and the intervals between the associated quantum jumps can be sorted according to the corresponding four different types ($mm')$:
\begin{itemize}[label={}]
  \item (00):\ \ $\ket{2}\rightarrow \ket{0}$ after $\ket{2}\rightarrow \ket{0}$
  \item (10):\ \  $\ket{2}\rightarrow \ket{0}$ after $\ket{2}\rightarrow \ket{1}$
  \item (01):\ \  $\ket{2}\rightarrow \ket{1}$ after $\ket{2}\rightarrow \ket{0}$
  \item (11):\ \  $\ket{2}\rightarrow \ket{1}$ after $\ket{2}\rightarrow \ket{1}$
\end{itemize}

Most physical systems are prone to dephasing, e.g., due to fluctuating magnetic fields, and we model this by introducing a decoherence term in the master equations (\ref{eq:master0}, \ref{eq:mastertilde}, \ref{eq:mastertildeeta}) corresponding to the operator $\hat{C}_D=\sqrt{\gamma}(\ket{0}\bra{0}-\ket{1}\bra{1}+\ket{2}\bra{2})$. The effect of this is to flip the sign of the $\ket{1}$ amplitude relative to those of the two other states with a rate $\gamma$.

\begin{figure}
\centering
\includegraphics[trim=0 0 0 0,width=0.9\columnwidth]{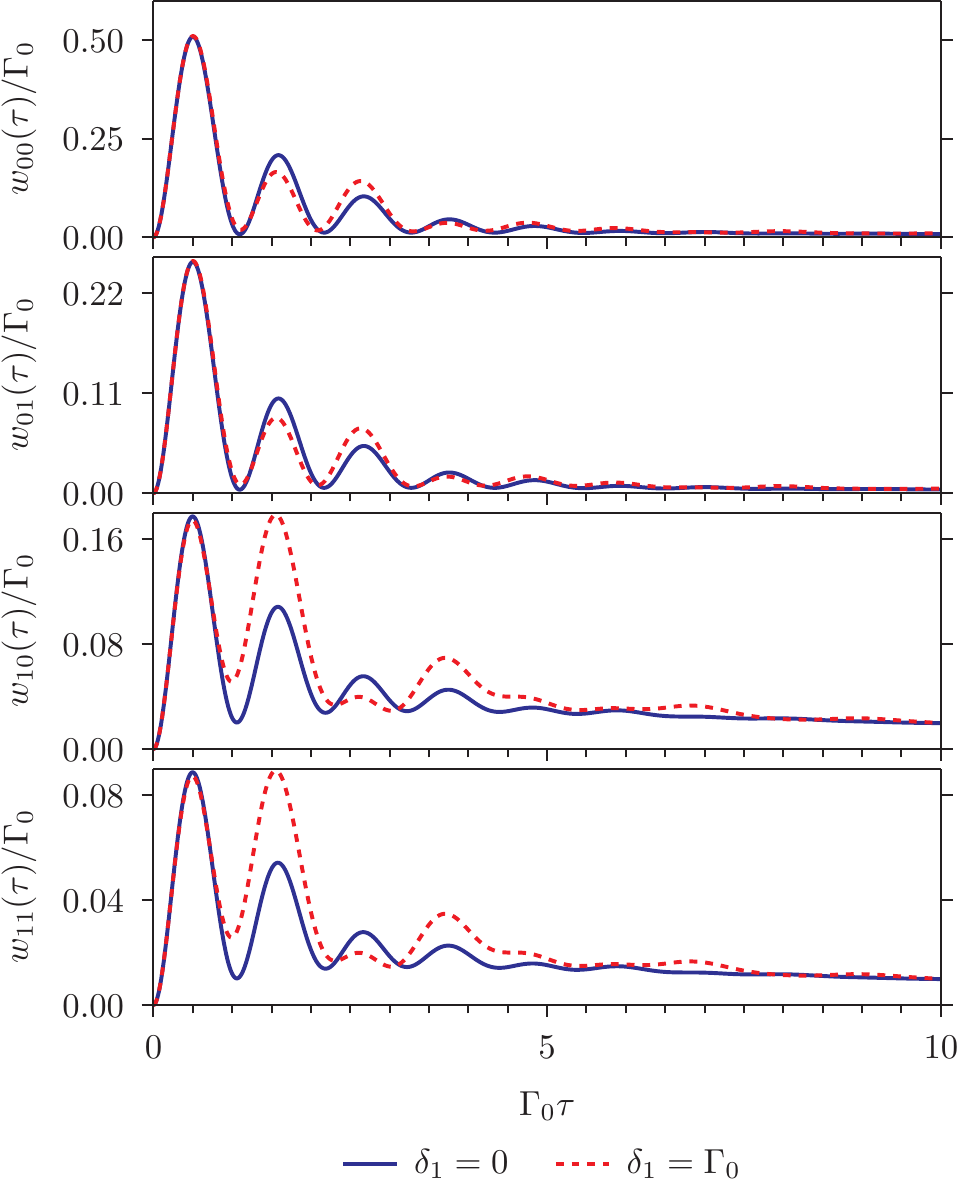}
\caption{\textsl{(Color online) Delay functions for each of the relevant interval types in a $\Lambda$-type system, calculated for
$\Omega_0 = 5\Gamma_0$,
$\Omega_1 = 3\Gamma_0$,
$\delta_0 = 0$,
$\Gamma_{1} = 0.5\Gamma_{0}$, and a ground state dephasing rate $\gamma=0.1\Gamma_0$. The blue, solid lines are for the resonant case  $\delta_1=0$, and the red, dotted lines are for the detuned case, $\delta_1=\Gamma_{0}$.
}}
\label{fig:w}
\end{figure}

In \fref{fig:w}, we show two examples of the four delay functions $w_{mm'}(\tau)$ for the $\Lambda$-system assuming perfect detection in both channels (physical parameters are given in the figure caption).

For resonant coupling  on both transitions (blue, solid lines) all four waiting time distributions resemble those of a two level system, (see \cite{delay}). For finite detuning (red dashed lines) of the $|1\rangle \leftrightarrow |2\rangle$ transition, the waiting time distribution functions after decay into $\ket{0}$ largely maintain the same form, while, after decay into $\ket{1}$ the distributions reflect the off-resonant $\ket{1}\rightarrow\ket{2}$ excitation process.

\begin{figure}
\centering
\includegraphics[width=0.9\columnwidth]{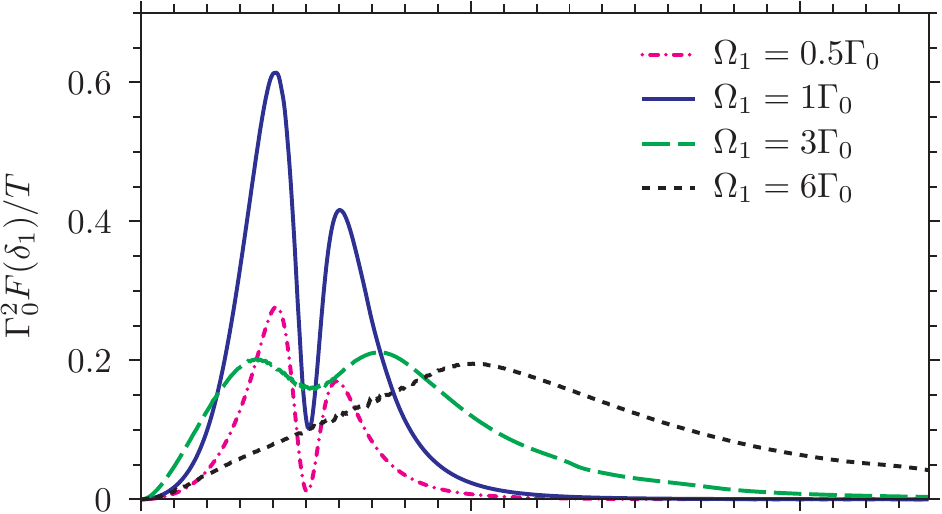}
\includegraphics[width=0.9\columnwidth]{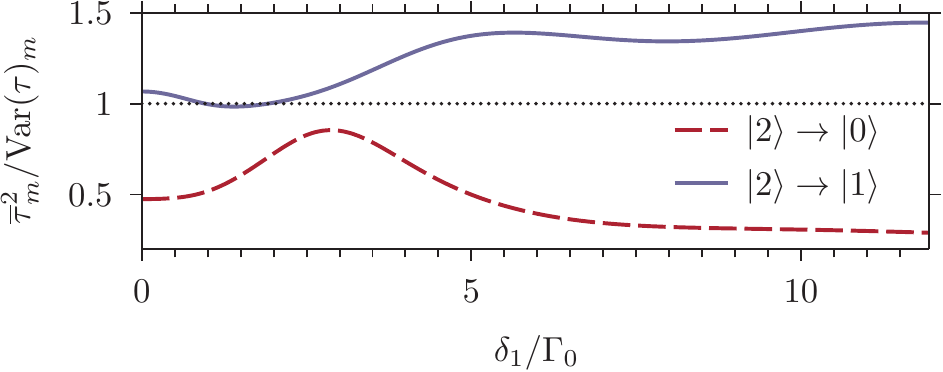}
\caption{\textsl{(Color online) Upper panel: The Fisher information per unit time for estimation of the laser-atom detuning $\delta_1$ by photon counting. Results are shown for different values of the Rabi frequency, from weak  $\Omega_1=0.5\Gamma_{0}$ to strong $\Omega_1=6\Gamma_{0}$, and the other parameters are
$\Omega_0 = 5\Gamma_0$,
$\delta_0 = 0$,
$\Gamma_{1} = 0.5\Gamma_0$, and $\gamma=0.1\Gamma_0$. For $\delta_0=0$, all statistical properties of the counting signal, and hence the Fisher information, are even functions of $\delta_1$.
Lower panel: The ratio $\overline{\tau}^2_m/\textrm{Var}(\tau)_m$ for the waiting times in the two channels as function of $\delta_1$. We assume $\Omega_1=3\Gamma_0$, while the remaining parameters are as in the upper panel.
}}
\label{fig:F_delta_s} 
\end{figure}

In \fref{fig:F_delta_s} we show in the upper panel the Fisher information divided by the probing time for the estimation of the detuning $\theta=\delta_1$ for different values of the laser Rabi frequency $\Omega_1$.
For $\delta_0=0$, all statistical properties of the counting signal are even functions of $\delta_1$ and, as witnessed by the vanishing Fisher information, we are not able to distinguish values of $\delta_1$ close to $\delta_1=0$.
At finite detuning, we obtain the highest Fisher information for $\Omega_1\sim\Gamma_0$. For weak driving ($\Omega_1=0.5\Gamma_0$) the $\ket{2}\leftrightarrow\ket{1}$ laser is a small perturbation in the Hamiltonian (\ref{hamilton lambda}), and the absorption spectrum is characterized by resonances at $\delta_1=\pm\Omega_0/2$, AC-Stark shifted by the strong $\ket{2}\leftrightarrow\ket{0}$ coupling laser. At resonances, the gradient of $\overline{n}_{mm'}(\tau,\delta_1)$ vanishes, and as seen from the distinct dip in the Fisher information our ability to discern different values of the detuning here vanishes in the limit $\Omega_1\rightarrow 0$.

In the lower panel of \fref{fig:F_delta_s} we show the ratio $\overline{\tau}^2_m/\textrm{Var}(\tau)_m$ for the two channels as function of $\delta_1$ and for $\Omega_1=3\Gamma_0$. According to \eqref{eq:V_m} the distribution of $N_m$ is sub-Possionian for values of this ratio larger than unity which occur for counts in the $\ket{2}\rightarrow\ket{1}$-channel for almost all values of $\delta_1$, and super-Possionian for values smaller than unity which occur in the $\ket{2}\rightarrow\ket{0}$-channel for all values of $\delta_1$, given the remaining parameters used in this example.

Let us also investigate the parameter estimation sensitivity for a system with multiple decay channels of which only one is being observed. This situation occurs, \textit{e.g.}, in solid state emitters, which may relax both optically and by non-radiative coupling to the host material, and in the case of atoms which decay by emission of light in very different wave length regions. To describe this situation, we introduce hypothetical observers, Alice and Bob, holding only partial detection records. Alice has a perfect detector that monitors only the $\ket{2}\rightarrow\ket{0}$ channel. Her record of waiting times must then be matched to the distribution $w_{00}(\tau)$ found from \eqref{eq:mastertildeeta}, solved for the initial state $\ket{0}$ with $\eta_{0}=1$ and with $\eta_{1}=0$.
Bob, on the other hand, monitors the $\ket{2}\rightarrow\ket{1}$ channel only, and his record of waiting times must be matched to the distribution $w_{11}(\tau)$ found from \eqref{eq:mastertildeeta} solved for the initial state $\ket{1}$ with $\eta_{0}=0$ and $\eta_{1}=1$. The middle time-line in Fig. 4(a) illustrates a full detection record while the upper (lower) line shows the detection record of Alice (Bob).
\begin{figure}
\subfigure[]{
\centering
\includegraphics[angle=0, width=0.9\columnwidth]{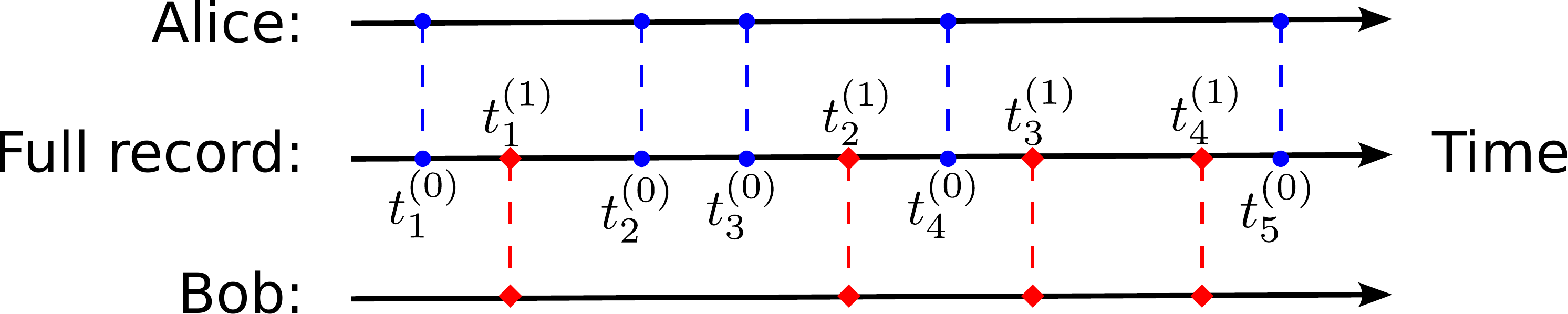}
\label{fig:records1}}
\subfigure[]{
\centering \includegraphics[angle=0, width=0.9\columnwidth]{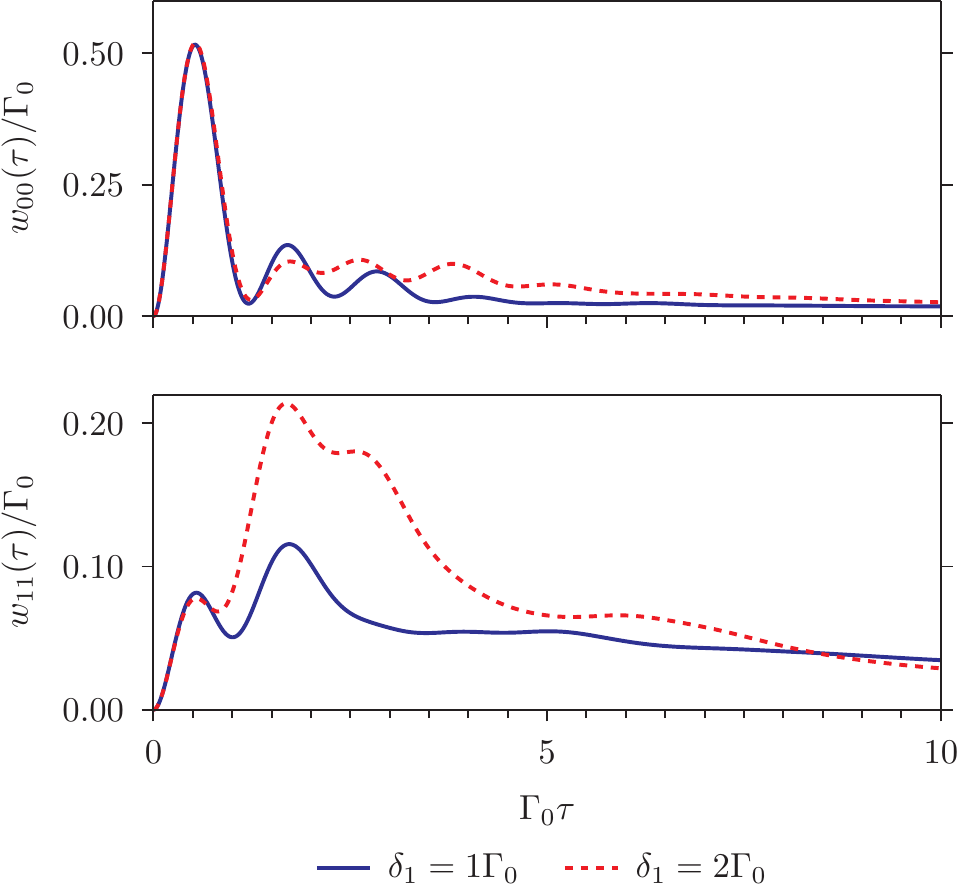}
\label{fig:AB_hist}}
\centering
\caption{\textsl{(Color online) (a) Schematic illustration of the detection records of Alice and Bob (see text) and a full detection record. The blue dots at times $t_i^{(0)}$ are emissions in the channel $\ket{2}\rightarrow\ket{0}$ monitored by Alice. The red diamonds at times $t_i^{(1)}$ are emissions in the channel $\ket{2}\rightarrow\ket{1}$ monitored by Bob. The observers do not see photons from the other channel. The full record holds information on all emission events.
(b) the waiting time distributions for the measurement records obtained by Alice (upper panel), monitoring only the channel $|2\rangle \rightarrow |0\rangle$,  and by Bob (lower panel), monitoring only the channel $|2\rangle \rightarrow |1\rangle$.
These are calculated for the parameter values
$\Omega_0 = 5\Gamma_0$,
$\Omega_1 = 2\Gamma_0$,
$\delta_0 = 0$,
$\Gamma_{1} = \Gamma_{0}$, and $\gamma=0.1\Gamma_0$ and shown for $\delta_1=\Gamma_0$ (blue, solid lines) and $\delta_1=2\Gamma_0$ (red, dashed lines) respectively.}}
\end{figure}

\begin{figure}
\centering
\includegraphics[width=0.9\columnwidth]{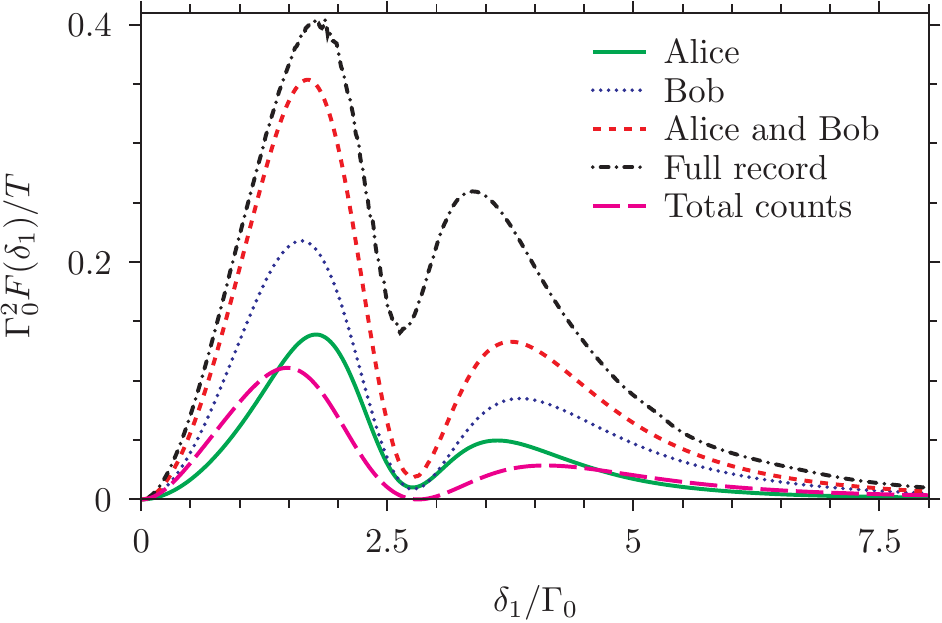}
\caption{\textsl{(Color online)
The Fisher information per time for estimation of the laser-atom detuning $\delta_1$ in a $\Lambda$-type system by photon counting by Alice (green, solid line) and Bob (blue, dotted line), and by use of their combined records of waiting times (red, short-dashed line). The Fisher information from the complete detection record of both channels is shown as the dashed-dotted black curve, while the sensitivity obtained by only utilizing the total photon count (\eqref{mean}) is shown by the purple, dashed curve.
The results are calculated for the parameters
$\Omega_0 = 5\Gamma_0$,
$\Omega_1 = 2\Gamma_0$,
$\delta_0 = 0$,
$\Gamma_{1} = 1$, and $\gamma=0.1\Gamma_0$.}}
\label{fig:ABCD} 
\end{figure}

In Fig. 4(b), we show the waiting time distributions for two values of the detuning $\delta_1$ (other physical parameters are given in the figure caption).
The achievements of optimal frequency estimation strategies based on the individual records of Alice and Bob are given by the Fisher information Eqs.(\ref{F},\ref{eq:F_Nfinal}), where the sum only has one term, $(mm')=(00)$ for Alice and $(mm')=(11)$ for Bob.
Combining their records of waiting times, however, Alice and Bob may achieve a higher level of sensitivity. The Fisher information is then the sum of the individual Fisher informations according to Eqs.(\ref{F},\ref{eq:F_Nfinal}). We show in \fref{fig:ABCD} the Fisher information per time for estimation of $\delta_1$ by the separate records of Alice and Bob and by combining their registered distribution of waiting times. In Fig. 4(b), we observe that the delay function connected to the channel $\ket{2}\rightarrow \ket{0}$ is less sensitive to changes in detuning than the one pertaining to the $\ket{2}\rightarrow \ket{1}$ channel. This explains why Bob outperforms Alice at estimating the value of $\delta_1$.


The Fisher information for the full detection record
(dash-dotted line in \fref{fig:ABCD}) is higher than that of Alice
and Bob, even when they combine their waiting time
records. This is because it makes use of all detection events and for example
recognizes the first interval  in Alice’s record in Fig. 4(a) as two subsequent
$(mm’)=(01)$ and $(10)$ intervals rather than a single $(00)$ interval.


Consider, finally, an observer who has only access to the total, accumulated photon count. For a general multi-channel emitter the mean photo current in the asymptotic limit is $\overline{N}/T=\sum_m \textrm{Tr}(\hat{C}_{m}^\dagger\hat{C}_{m}\rho^{st}(\tau))$.  For general counting statistics, we have $\Delta N = \sqrt{V}$. This implies an uncertainty on $\theta$ given by
$\Delta \theta=(\partial N/\partial\theta)^{-1}\sqrt{V}$, \textit{i.e.}, for detuning estimation in our $\Lambda$-atom,
\begin{align}
\frac{(\Delta\delta_1)^{-2}}{T}
=
\frac{(\Gamma_0+\Gamma_1)^2}{\Gamma_{0}\frac{\textrm{Var}(\tau)_0}{\overline{\tau}_0^2}+\Gamma_{1}\frac{\textrm{Var}(\tau)_1}{\overline{\tau}_1^2}} 
\frac{(\partial \rho_{22}^{st}/\partial \delta_1)^2}{\rho_{22}^{st}},
\label{mean}
\end{align}
where we have used \eqref{eq:V_m} and $V=V_0+V_1$.
By \eqref{eq:CRB} this can be directly compared to the Fisher information per time, and the result of \eqref{mean} is included as the purple, long-dashed curve in \fref{fig:ABCD}.
As expected, parameter estimates obtained from the full record and from the combined waiting time records of Alice and Bob achieve higher sensitivity on the whole detuning range.
%

\section{Conclusion}
The full photo detection record of a quantum emitter contains more information about its dynamics than the mean signal. In this article, we have formulated a theory that quantifies this by calculating the Cramér-Rao sensitivity limit for multi-channel quantum light emitters:
The information in the full photo detection record may be represented as waiting time distributions for which Eqs. (17,18) provide theoretical results, and which, by \eqref{eq-non-poisson}, supply the fundamental sensitivity limit \eqref{eq:CRB}.
This optimal limit may be achieved via the linear estimator \eqref{general:S} or by a maximum likelihood estimate \cite{likelihood}.
%
We exemplified the theory by the estimation of a detuning parameter in a driven $\Lambda$-type system with two distinct decay channels.

Our theory assumes an ergodic emitter, \textit{i.e.}, the system has a steady state which does not depend on the initial state of the system and which is not a dark state, such that the amount of accumulated data grows linearly with time. We also assumed that the decay of the system always feeds the same discrete set of final states, so that the data record can be analysed by a finite number of waiting time functions.  Both the ergodicity assumption and the restriction to a finite number of final states are technical conditions for our method to apply, while our underlying Bayesian description is readily applied and several of the concepts introduced in this paper can be modified to account for the sensitivity limit in more general cases.
\\\\
\section{Acknowledgments}
The authors acknowledge financial support from the Villum Foundation and helpful comments on the manuscript from   C. K. Andersen,  D. D. Bhaktavatsala Rao, P. Haikka, X. Qing, M. C. Tichy, and A. J. C. Wade.


\begin{thebibliography}{15}%
\makeatletter
\providecommand \@ifxundefined [1]{%
 \@ifx{#1\undefined}
}%
\providecommand \@ifnum [1]{%
 \ifnum #1\expandafter \@firstoftwo
 \else \expandafter \@secondoftwo
 \fi
}%
\providecommand \@ifx [1]{%
 \ifx #1\expandafter \@firstoftwo
 \else \expandafter \@secondoftwo
 \fi
}%
\providecommand \natexlab [1]{#1}%
\providecommand \enquote  [1]{``#1''}%
\providecommand \bibnamefont  [1]{#1}%
\providecommand \bibfnamefont [1]{#1}%
\providecommand \citenamefont [1]{#1}%
\providecommand \href@noop [0]{\@secondoftwo}%
\providecommand \href [0]{\begingroup \@sanitize@url \@href}%
\providecommand \@href[1]{\@@startlink{#1}\@@href}%
\providecommand \@@href[1]{\endgroup#1\@@endlink}%
\providecommand \@sanitize@url [0]{\catcode `\\12\catcode `\$12\catcode
  `\&12\catcode `\#12\catcode `\^12\catcode `\_12\catcode `\%12\relax}%
\providecommand \@@startlink[1]{}%
\providecommand \@@endlink[0]{}%
\providecommand \url  [0]{\begingroup\@sanitize@url \@url }%
\providecommand \@url [1]{\endgroup\@href {#1}{\urlprefix }}%
\providecommand \urlprefix  [0]{URL }%
\providecommand \Eprint [0]{\href }%
\providecommand \doibase [0]{http://dx.doi.org/}%
\providecommand \selectlanguage [0]{\@gobble}%
\providecommand \bibinfo  [0]{\@secondoftwo}%
\providecommand \bibfield  [0]{\@secondoftwo}%
\providecommand \translation [1]{[#1]}%
\providecommand \BibitemOpen [0]{}%
\providecommand \bibitemStop [0]{}%
\providecommand \bibitemNoStop [0]{.\EOS\space}%
\providecommand \EOS [0]{\spacefactor3000\relax}%
\providecommand \BibitemShut  [1]{\csname bibitem#1\endcsname}%
\let\auto@bib@innerbib\@empty
\bibitem [{\citenamefont {Giovannetti}\ \emph {et~al.}(2004)\citenamefont
  {Giovannetti}, \citenamefont {Lloyd},\ and\ \citenamefont
  {Maccone}}]{Giovannetti19112004}%
  \BibitemOpen
  \bibfield  {author} {\bibinfo {author} {\bibfnamefont {V.}~\bibnamefont
  {Giovannetti}}, \bibinfo {author} {\bibfnamefont {S.}~\bibnamefont {Lloyd}},
  \ and\ \bibinfo {author} {\bibfnamefont {L.}~\bibnamefont {Maccone}},\ }\href
  {\doibase 10.1126/science.1104149} {\bibfield  {journal} {\bibinfo  {journal}
  {Science}\ }\textbf {\bibinfo {volume} {306}},\ \bibinfo {pages} {1330}
  (\bibinfo {year} {2004})}\BibitemShut {NoStop}%
\bibitem [{\citenamefont {T\'oth}\ and\ \citenamefont
  {Apellaniz}(2014)}]{review}%
  \BibitemOpen
  \bibfield  {author} {\bibinfo {author} {\bibfnamefont {G.}~\bibnamefont
  {T\'oth}}\ and\ \bibinfo {author} {\bibfnamefont {I.}~\bibnamefont
  {Apellaniz}},\ }\href {http://stacks.iop.org/1751-8121/47/i=42/a=424006}
  {\bibfield  {journal} {\bibinfo  {journal} {J. Phys. A: Math. Theor.}\
  }\textbf {\bibinfo {volume} {47}},\ \bibinfo {pages} {424006} (\bibinfo
  {year} {2014})}\BibitemShut {NoStop}%
\bibitem [{\citenamefont {Mabuchi}(1996)}]{Mabuchi1996}%
  \BibitemOpen
  \bibfield  {author} {\bibinfo {author} {\bibfnamefont {H.}~\bibnamefont
  {Mabuchi}},\ }\href {http://stacks.iop.org/1355-5111/8/i=6/a=002} {\bibfield
  {journal} {\bibinfo  {journal} {Quantum Semiclass. Opt.}\ }\textbf {\bibinfo
  {volume} {8}},\ \bibinfo {pages} {1103} (\bibinfo {year} {1996})}\BibitemShut
  {NoStop}%
\bibitem [{\citenamefont {Gambetta}\ and\ \citenamefont
  {Wiseman}(2001)}]{PhysRevA.64.042105}%
  \BibitemOpen
  \bibfield  {author} {\bibinfo {author} {\bibfnamefont {J.}~\bibnamefont
  {Gambetta}}\ and\ \bibinfo {author} {\bibfnamefont {H.~M.}\ \bibnamefont
  {Wiseman}},\ }\href {\doibase 10.1103/PhysRevA.64.042105} {\bibfield
  {journal} {\bibinfo  {journal} {Phys. Rev. A}\ }\textbf {\bibinfo {volume}
  {64}},\ \bibinfo {pages} {042105} (\bibinfo {year} {2001})}\BibitemShut
  {NoStop}%
\bibitem [{\citenamefont {Gammelmark}\ and\ \citenamefont
  {M\o{}lmer}(2013)}]{likelihood}%
  \BibitemOpen
  \bibfield  {author} {\bibinfo {author} {\bibfnamefont {S.}~\bibnamefont
  {Gammelmark}}\ and\ \bibinfo {author} {\bibfnamefont {K.}~\bibnamefont
  {M\o{}lmer}},\ }\href {\doibase 10.1103/PhysRevA.87.032115} {\bibfield
  {journal} {\bibinfo  {journal} {Phys. Rev. A}\ }\textbf {\bibinfo {volume}
  {87}},\ \bibinfo {pages} {032115} (\bibinfo {year} {2013})}\BibitemShut
  {NoStop}%
\bibitem [{\citenamefont {{Cram\'{e}r}}(1954)}]{Cramer}%
  \BibitemOpen
  \bibfield  {author} {\bibinfo {author} {\bibfnamefont {H.}~\bibnamefont
  {{Cram\'{e}r}}},\ }\bibfield  {booktitle} {\emph {\bibinfo {booktitle}
  {{Mathematical methods of statistics}}},\ }\href
  {http://stacks.iop.org/1367-2630/13/i=5/a=053035} {\bibfield  {journal}
  {\bibinfo  {journal} {Princeton mathematical series No. 9 (Princeton
  University Press, Princeton)}\ } (\bibinfo {year} {1954})}\BibitemShut
  {NoStop}%
  \bibitem [{\citenamefont {{Cox}}(1970)}]{Cox}%
  \BibitemOpen
  \bibfield  {author} {\bibinfo {author} {\bibfnamefont {D.}~\bibnamefont
  {{Cox}}},\ }\bibfield  {booktitle} {\emph {\bibinfo {booktitle}
  {{Renewal Theory}}},\ }\href
  {http://http://alg.csie.ncnu.edu.tw/~ykshieh/b1.pdf} {\bibfield  {journal}
  {\bibinfo  {journal} {(Methuen \& Co., London)}\ } (\bibinfo {year} {1970})}\BibitemShut
  {NoStop}%
  \bibitem [{\citenamefont {{Feller}}(1950)}]{Feller}%
  \BibitemOpen
  \bibfield  {author} {\bibinfo {author} {\bibfnamefont {W.}~\bibnamefont
  {{Feller}}},\ }\bibfield  {booktitle} {\emph {\bibinfo {booktitle}
  {{An Introduction to Probability Theory and Its Applications, Third Edition}}},\ }\href
  {http://http://alg.csie.ncnu.edu.tw/~ykshieh/b1.pdf} {\bibfield  {journal}
  {\bibinfo  {journal} {Chapter 9 (John Wiley \& Sons, Inc.)}\ } (\bibinfo {year} {1968})}\BibitemShut
  {NoStop}%
\bibitem [{\citenamefont {Kiilerich}\ and\ \citenamefont
  {M\o{}lmer}(2014)}]{delay}%
  \BibitemOpen
  \bibfield  {author} {\bibinfo {author} {\bibfnamefont {A.~H.}\ \bibnamefont
  {Kiilerich}}\ and\ \bibinfo {author} {\bibfnamefont {K.}~\bibnamefont
  {M\o{}lmer}},\ }\href {\doibase 10.1103/PhysRevA.89.052110} {\bibfield
  {journal} {\bibinfo  {journal} {Phys. Rev. A}\ }\textbf {\bibinfo {volume}
  {89}},\ \bibinfo {pages} {052110} (\bibinfo {year} {2014})}\BibitemShut
  {NoStop}%
\bibitem [{\citenamefont {Delaubert}\ \emph {et~al.}(2008)\citenamefont
  {Delaubert}, \citenamefont {Treps}, \citenamefont {Fabre}, \citenamefont
  {Bachor},\ and\ \citenamefont {Réfrégier}}]{CRB}%
  \BibitemOpen
  \bibfield  {author} {\bibinfo {author} {\bibfnamefont {V.}~\bibnamefont
  {Delaubert}}, \bibinfo {author} {\bibfnamefont {N.}~\bibnamefont {Treps}},
  \bibinfo {author} {\bibfnamefont {C.}~\bibnamefont {Fabre}}, \bibinfo
  {author} {\bibfnamefont {H.~A.}\ \bibnamefont {Bachor}}, \ and\ \bibinfo
  {author} {\bibfnamefont {P.}~\bibnamefont {Réfrégier}},\ }\href
  {http://stacks.iop.org/0295-5075/81/i=4/a=44001} {\bibfield  {journal}
  {\bibinfo  {journal} {EPL}\ }\textbf {\bibinfo {volume} {81}},\ \bibinfo
  {pages} {44001} (\bibinfo {year} {2008})}\BibitemShut {NoStop}%
\bibitem [{\citenamefont {Negretti}\ \emph {et~al.}(2008)\citenamefont
  {Negretti}, \citenamefont {Henkel},\ and\ \citenamefont
  {M\o{}lmer}}]{mereCRB}%
  \BibitemOpen
  \bibfield  {author} {\bibinfo {author} {\bibfnamefont {A.}~\bibnamefont
  {Negretti}}, \bibinfo {author} {\bibfnamefont {C.}~\bibnamefont {Henkel}}, \
  and\ \bibinfo {author} {\bibfnamefont {K.}~\bibnamefont {M\o{}lmer}},\ }\href
  {\doibase 10.1103/PhysRevA.78.023630} {\bibfield  {journal} {\bibinfo
  {journal} {Phys. Rev. A}\ }\textbf {\bibinfo {volume} {78}},\ \bibinfo
  {pages} {023630} (\bibinfo {year} {2008})}\BibitemShut {NoStop}%
\bibitem [{\citenamefont {Paul}(1982)}]{RevModPhys.54.1061}%
  \BibitemOpen
  \bibfield  {author} {\bibinfo {author} {\bibfnamefont {H.}~\bibnamefont
  {Paul}},\ }\href {\doibase 10.1103/RevModPhys.54.1061} {\bibfield  {journal}
  {\bibinfo  {journal} {Rev. Mod. Phys.}\ }\textbf {\bibinfo {volume} {54}},\
  \bibinfo {pages} {1061} (\bibinfo {year} {1982})}\BibitemShut {NoStop}%
\bibitem [{\citenamefont {Wiseman}\ and\ \citenamefont {Milburn}(2010)}]{WM}%
  \BibitemOpen
  \bibfield  {author} {\bibinfo {author} {\bibfnamefont {H.~M.}\ \bibnamefont
  {Wiseman}}\ and\ \bibinfo {author} {\bibfnamefont {G.~J.}\ \bibnamefont
  {Milburn}},\ }\bibfield  {booktitle} {\emph {\bibinfo {booktitle} {{Quantum
  Measurement and Control}}},\ }\href@noop {} {\bibfield  {journal} {\bibinfo
  {journal} {(Cambridge University Press, Cambridge)}\ } (\bibinfo {year}
  {2010})}\BibitemShut {NoStop}%
\bibitem [{\citenamefont {{Carmichael}}(1993)}]{CARMICHAEL}%
  \BibitemOpen
  \bibfield  {author} {\bibinfo {author} {\bibfnamefont {H.}~\bibnamefont
  {{Carmichael}}},\ }\bibfield  {booktitle} {\emph {\bibinfo {booktitle} {{An
  Open Systems Approach to Quantum Optics, Lectures Presented at the
  Universit\'{e} Libre de Bruxelles October 28 to November 4, 1991}}},\ }\href
  {http://stacks.iop.org/1367-2630/13/i=5/a=053035} {\bibfield  {journal}
  {\bibinfo  {journal} {(Springer Berlin Heidelberg)}\ } (\bibinfo {year}
  {1993})}\BibitemShut {NoStop}%
\bibitem [{\citenamefont {Dalibard}\ \emph {et~al.}(1992)\citenamefont
  {Dalibard}, \citenamefont {Castin},\ and\ \citenamefont {M\o{}lmer}}]{MCWF}%
  \BibitemOpen
  \bibfield  {author} {\bibinfo {author} {\bibfnamefont {J.}~\bibnamefont
  {Dalibard}}, \bibinfo {author} {\bibfnamefont {Y.}~\bibnamefont {Castin}}, \
  and\ \bibinfo {author} {\bibfnamefont {K.}~\bibnamefont {M\o{}lmer}},\ }\href
  {\doibase 10.1103/PhysRevLett.68.580} {\bibfield  {journal} {\bibinfo
  {journal} {Phys. Rev. Lett.}\ }\textbf {\bibinfo {volume} {68}},\ \bibinfo
  {pages} {580} (\bibinfo {year} {1992})}\BibitemShut {NoStop}%
\bibitem [{\citenamefont {{Gammelmark}}\ and\ \citenamefont
  {{Mølmer}}(2011)}]{gammelmark}%
  \BibitemOpen
  \bibfield  {author} {\bibinfo {author} {\bibfnamefont {S.}~\bibnamefont
  {{Gammelmark}}}\ and\ \bibinfo {author} {\bibfnamefont {K.}~\bibnamefont
  {{Mølmer}}},\ }\href {http://stacks.iop.org/1367-2630/13/i=5/a=053035}
  {\bibfield  {journal} {\bibinfo  {journal} {New J. Phys}\ }\textbf {\bibinfo
  {volume} {13}},\ \bibinfo {pages} {053035} (\bibinfo {year}
  {2011})}\BibitemShut {NoStop}%
\bibitem [{\citenamefont {Gill}\ and\ \citenamefont {Guta}(2013)}]{GutaGill}%
  \BibitemOpen
  \bibfield  {author} {\bibinfo {author} {\bibfnamefont {R.~D.}\ \bibnamefont
  {Gill}}\ and\ \bibinfo {author} {\bibfnamefont {M.~I.}\ \bibnamefont
  {Guta}},\ }\bibfield  {booktitle} {\emph {\bibinfo {booktitle} {On asymptotic
  quantum statistical inference in From Probability to Statistics and Back:
  High-Dimensional Models and Processes -- A Festschrift in Honor of Jon A.
  Wellner}},\ }\href@noop {} {\bibfield  {journal} {\bibinfo  {journal}
  {(Institute of Mathematical Statistics, Beachwood)}\ } (\bibinfo {year}
  {2013})}\BibitemShut {NoStop}%
\end{thebibliography}
%

\end{document}